\documentclass[cameraready]{Interspeech}

\title{SphereVBx: Spherical Variational Bayes Clustering for Simplified EEND-VC Diarization} 

\author[affiliation={1}, orcid=0009-0000-9659-3004]{Petr}{Pálka}
\author[affiliation={1}, orcid=0000-0001-5390-8520]{Jiangyu}{Han}
\author[affiliation={1}, orcid=0000-0003-0760-8572]{Prachi}{Singh}
\author[affiliation={2}, orcid=0000-0002-5175-7834]{Marc}{Delcroix}
\author[affiliation={2}, orcid=0000-0002-1130-5059]{Naohiro}{Tawara}
\author[affiliation={1}, orcid=0000-0002-4951-5908]{\ \ \ \ \ Lukáš}{Burget}

\address{
    $^1$ Brno University of Technology,	Czechia, 
    $^2$ NTT, Inc.,	Japan
}

\email{ipalka@fit.vutbr.cz}

\keywords{Speaker diarization, EEND-VC, PSDA, VBx} 

\usepackage{comment}
\usepackage{amsmath}
\usepackage{amssymb}
\usepackage{soul}

\usepackage{makecell} 
\usepackage{array} 

\usepackage{threeparttable}
\newcolumntype{C}[1]{>{\centering\arraybackslash}b{#1}} 
\newcolumntype{M}[1]{>{\centering\arraybackslash}m{#1}}

 \usepackage{graphicx}%
 \usepackage{subcaption}
\usepackage{tikz}
\usetikzlibrary{arrows.meta,positioning,calc,fit,decorations.pathreplacing}

\begin{document}

\maketitle
\vspace{-2mm}

\begin{abstract}
    We propose SphereVBx, a Bayesian clustering framework for hyperspherical embeddings based on Toroidal Probabilistic Spherical Discriminant Analysis (T-PSDA). The method follows the variational Bayesian formulation of VBx while replacing the Gaussian Probabilistic Linear Discriminant Analysis (PLDA) backend with T-PSDA, resulting in variational inference in a mixture of von~Mises–Fisher distributions. We apply SphereVBx to speaker diarization and in particular to the end-to-end neural diarization with vector clustering (EEND-VC) framework. A parameter-free variant, denoted SphereVBx-PF, corresponds to a spherical similarity model closely related to cosine scoring and does not require pretrained backend parameters. Experiments on multiple diarization benchmarks show that SphereVBx improves clustering accuracy in cascaded diarization pipelines and achieves comparable or better performance in the EEND-VC framework while significantly simplifying its clustering stage.

\end{abstract}

\section{Introduction}
Many modern representation learning systems produce embeddings that are normalized to lie on the unit hypersphere. This is common in tasks such as speaker recognition \cite{xiang2019margin} and face recognition \cite{deng2019arcface}. Although cosine similarity is widely used to compare such embeddings, probabilistic clustering methods that explicitly model their hyperspherical geometry are less commonly used in practice \cite{banerjee2005clustering, gopal2014mises, dubey2018robust, sholokhov2023}.

In this work, we introduce \emph{SphereVBx}, a Bayesian clustering framework for hyperspherical embeddings based on mixtures of von~Mises–Fisher (vMF) distributions. The~method builds on the Variational Bayes clustering framework of VBx~\cite{diez19_interspeech,landini2022bayesian}, which has become a strong baseline for speaker diarization. Standard VBx models speaker embeddings using a Gaussian Probabilistic Linear Discriminant Analysis (PLDA) \cite{brummer10_odyssey} backend. However, PLDA does not fully reflect the properties of modern embeddings that are typically length-normalized and trained with angular-margin objectives \cite{deng2019arcface}. In practice, simple cosine similarity scoring is often observed to outperform PLDA for such embeddings, suggesting that spherical similarity measures may be more appropriate \cite{PSDA_brummer22_interspeech}. SphereVBx replaces the PLDA model used in VBx with Toroidal Probabilistic Spherical Discriminant Analysis (T-PSDA)~\cite{TPSDA_10095580}, resulting in variational inference in a Bayesian mixture of \mbox{von~Mises–Fisher distributions}.

We apply SphereVBx to speaker diarization, where speaker embeddings extracted from short audio segments are clustered to assign speaker labels. We evaluate SphereVBx in both traditional cascaded~\cite{landini2022bayesian,ryant21_interspeech} and end-to-end~\cite{bredin23_interspeech} diarization pipelines.
In particular, we focus on the end-to-end neural diarization with vector clustering (EEND-VC) framework~\cite{kinoshita2021integrating,advances_kino}, which combines a local EEND model that estimates speaker activity within short overlapping windows with a global clustering stage that links these local speaker instances across the full recording.

Recent advances \cite{bredin23_interspeech, palka2025vbx, liao2026dual} in the EEND-VC setting rely on several heuristic steps, including filtering out embeddings deemed unreliable before clustering and cosine-similarity-based reassignment after clustering. SphereVBx provides a principled probabilistic alternative that operates directly on the embeddings and eliminates the need for these ad-hoc steps. A special parameter setting yields a \emph{parameter-free SphereVBx} variant that does not require pretrained backend parameters and whose underlying similarity model is related to cosine scoring.

We further introduce reliability weights that reduce the contribution of unreliable embeddings to the clustering, and evaluate SphereVBx on eight standard diarization benchmarks. Our implementation is publicly available.\footnote{https://github.com/BUTSpeechFIT/DiariZen}

\vspace{-0.5mm}
\section{Background}
\subsection{PLDA-based VBx}  
\label{sec:PLDA_VBx}
VBx was introduced in~\cite{diez19_interspeech} as a Bayesian framework for clustering a sequence of speaker embeddings $\mathbf{X} = [\mathbf{x}_1, \dots, \mathbf{x}_T]$ extracted from short, overlapping speech segments. VBx models $\mathbf{X}$ using a Hidden Markov Model (HMM) in which each state corresponds to a speaker and transitions represent speaker changes. Conditioned on speaker state $s$, embeddings follow a Gaussian distribution $\mathcal{N}(\mathbf{x}_t \mid \mathbf{m}_s, \mathbf{\Sigma}_w)$, where $\mathbf{m}_s$ is the speaker-specific mean and $\mathbf{\Sigma}_w$ is a shared within-speaker covariance matrix.
The model is Bayesian in that the speaker means are treated as latent variables with prior distribution $\mathcal{N}(\mathbf{m}_s \mid \mathbf{m}, \mathbf{\Sigma}_b)$, where $\mathbf{m}$ is the global embedding mean and $\mathbf{\Sigma}_b$ the between-speaker covariance. This hierarchical Gaussian model corresponds to the standard two-covariance PLDA model~\cite{twO_covar_villalba11_interspeech}. In practice, the parameters $\mathbf{m}$, $\mathbf{\Sigma}_w$, and $\mathbf{\Sigma}_b$ are obtained from a PLDA model trained on a large corpus of speaker embeddings and kept fixed during diarization.

For efficient inference, embeddings are centered by subtracting $\mathbf{m}$ and linearly transformed so that $\mathbf{\Sigma}_w$ becomes identity and $\mathbf{\Sigma}_b$ is diagonalized. With this preprocessing, the model can be reparameterized~\cite{landini2022bayesian} as
\begin{align}
\label{eq:PLDA_P_W}
p(\mathbf{x}_t \mid \mathbf{y}_s) &= \mathcal{N}(\mathbf{x}_t \mid \mathbf{V}\mathbf{y}_s, \mathbf{I}), \\
\label{eq:PLDA_P_B}
p(\mathbf{y}_s) &= \mathcal{N}(\mathbf{y}_s \mid \mathbf{0}, \mathbf{I}),
\end{align}
where $\mathbf{y}_s$ is a latent variable capturing speaker variability and $\mathbf{V}$ maps this latent variable to the embedding space. This form is commonly used in VBx and will serve as a reference for the modifications introduced later.


In this work, we use a simplified variant where the HMM is replaced by a Gaussian Mixture Model (GMM), removing temporal modeling. This leads to faster inference and performs comparably to the full HMM formulation~\cite{klement2024discriminative}. 
Moreover, in the EEND-VC setting considered here, multiple embeddings are produced per local window, making temporal ordering less meaningful and favoring the GMM formulation.

Given $\mathbf{X}$, diarization amounts to inferring latent assignments $z_t$ that associate each embedding $\mathbf{x}_t$ with a GMM component (speaker). VBx performs this using Variational Bayes inference~\cite{landini2022bayesian}, which iteratively estimates approximate posterior distributions $q(\mathbf{y}_s)$ and $q(z_t=s)=\gamma_{ts}$. This Bayesian formulation enables VBx to reliably determine the effective number of speakers through Automatic Relevance Determination~\cite{bishop2006pattern}, automatically suppressing redundant mixture components as their weights converge toward zero.

\subsection{Toroidal Probabilistic Spherical Discriminant Analysis}
\label{subsec:tpsda}
Analogously to PLDA, T-PSDA~\cite{TPSDA_10095580} models speaker embeddings and enables computation of log-likelihood ratio scores for speaker verification. Unlike PLDA, which relies on Gaussian distributions, T-PSDA assumes a vMF distribution for embeddings on the unit hypersphere $\mathbb{S}^{D-1}$, better matching the geometry of modern embeddings that are typically length-normalized and trained with angular-margin objectives.
For a unit-norm vector $\mathbf{x} \in \mathbb{S}^{D-1}$, the vMF distribution is given by
$\mathcal{V}(\mathbf{x}\mid~\boldsymbol{\mu}, \kappa)
\propto
\exp(\kappa\,\boldsymbol{\mu}^\top \mathbf{x}),
$ where $\boldsymbol{\mu} \in \mathbb{S}^{D-1}$ denotes the mean direction and $\kappa$ is the concentration parameter.

T-PSDA assumes embeddings of speaker $s$ follow
\begin{align}
\label{eq:TPSDA_P_W}
p(\mathbf{x}_t \mid \mathbf{y}_s)
&= \mathcal{V}(\mathbf{x}_t \mid \mathbf{K}\mathbf{y}_s,\kappa_w),
\end{align}
where $\mathbf{y}_s \in \mathbb{S}^{d-1}$ is a speaker-specific direction, $\mathbf{K} \in \mathbb{R}^{D\times d}$ defines a $d$-dimensional speaker subspace whose columns form an orthonormal basis, and $\kappa_w$ controls within-speaker concentration. The speaker directions are further assumed to follow
\begin{align}
\label{eq:TPSDA_P_B}
p(\mathbf{y}_s) &= \mathcal{V}(\mathbf{y}_s \mid \mathbf{v},\kappa_b),
\end{align}
where $\mathbf{v}$ is the global mean direction and $\kappa_b$ controls between-speaker concentration. In these equations, we consider a simplified T-PSDA variant with a single speaker subspace and without channel factors, which was shown to perform well~\cite{TPSDA_10095580} and is used throughout this work.
Like PLDA, the model parameters can be trained using the Expectation-Maximization (EM) algorithm~\cite{TPSDA_10095580} on large collections of embeddings from many speakers.

When $d=D$, T-PSDA reduces to Probabilistic Spherical Discriminant Analysis (PSDA)~\cite{PSDA_brummer22_interspeech}. In the further special case $\kappa_b=0$ and $\kappa_w=1$, the resulting log-likelihood ratio scores for a pair of embeddings become a monotonic function of cosine similarity between the embeddings. This connection allows cosine similarity scoring—often observed to outperform PLDA for modern embeddings—to be incorporated into the probabilistic VBx framework, as described in the next section. In particular, this property is useful in the EEND-VC setting, where using cosine similarity for embedding re-assignment after VBx clustering was found very effective in the DiariZen system~\cite{palka2025vbx}.

\vspace{-2mm}
\section{SphereVBx} 
\label{sec:SphereVBx}
SphereVBx combines VBx (Section~\ref{sec:PLDA_VBx}) with the T-PSDA model (Section~\ref{subsec:tpsda}). In particular, the PLDA within- and between-speaker distributions \eqref{eq:PLDA_P_W} and \eqref{eq:PLDA_P_B} used in VBx are replaced by the corresponding T-PSDA distributions \eqref{eq:TPSDA_P_W} and \eqref{eq:TPSDA_P_B}. The remaining VB inference framework of VBx is retained.

As discussed in Section~\ref{subsec:tpsda}, when $d=D$, $\kappa_b=0$, and $\kappa_w=1$, the T-PSDA model underlying SphereVBx corresponds to a cosine-similarity scoring model and does not require any pretrained parameters. We refer to this configuration as parameter-free SphereVBx (SphereVBx-PF).

Due to space limitations, we summarize only the update equations that differ from the standard PLDA-based VBx. These updates are derived analogously to PLDA-based VBx~\cite{landini2022bayesian,diez2021vbx}, replacing the PLDA likelihood and prior distributions with their T-PSDA counterparts~\cite{TPSDA_10095580}.
VB inference alternates between estimating speaker assignments $q(z_t=s)$ and updating the variational posterior $q(\mathbf{y}_s)$. Due to the self-conjugacy of the vMF family, this posterior remains vMF,
$q(\mathbf{y}_s) =
\mathcal{V}\left(
\mathbf{y}_s
\middle|
\frac{\mathbf{a}_s}{\lVert\mathbf{a}_s\rVert},
\lVert\mathbf{a}_s\rVert
\right)$, where the vector $\mathbf{a}_s$ is updated as
\begin{equation}
\label{eq:speaker_models_update}
\mathbf{a}_s =
\kappa_b\,\mathbf{v} + \frac{F_A}{F_B}\,\kappa_w
\sum_t \gamma_{ts}\,\mathbf{K}^\top \mathbf{x}_t,
\end{equation}
where $F_A$ and $F_B$ are scaling hyperparameters introduced in VBx~\cite{landini2022bayesian}. Smaller values of $F_A$ compensate for overconfident posteriors caused by the independence assumption between embeddings, while larger values of $F_B$ encourage pruning of redundant speakers.

Here $\gamma_{ts}=q(z_t=s)$ denotes the posterior responsibility of assigning embedding $\mathbf{x}_t$ to speaker $s$. In our simplified formulation, which replaces the HMM of VBx with a Bayesian mixture model~\cite{klement2024discriminative,palka2025vbx}, the responsibilities are updated as
\begin{equation}
\label{eq:speaker_responsibilities_update}
\gamma_{ts} \propto \bar{p}(\mathbf{x}_t \mid s)\,\pi_s,
\end{equation}
where, analogously to VBx~\cite{diez2021vbx}, $\bar{p}(\mathbf{x}_t \mid s)$ is defined through the expected log-likelihood
\begin{equation}
\label{eq:expected_llh}
\begin{aligned}
\log \overline{p}(\mathbf{x}_t|s)
&\propto 
\mathbb{E}_{q(\mathbf{y}_s)}
\left[
F_A \log p(\mathbf{x}_t|\mathbf{y}_s)
\right] \\
&\propto 
F_A \kappa_w
(\mathbf{K}\hat{\boldsymbol{\mu}}_s)^{\top}\mathbf{x}_t,
\end{aligned}
\end{equation}
where the second line follows from evaluating the expectation under the vMF posterior $q(\mathbf{y}_s)$, whose mean is given by
$\hat{\boldsymbol{\mu}}_s=\mathbb{E}_{q(\mathbf{y}_s)}[\mathbf{y}_s]
=
\boldsymbol{\mu}_s
\frac{I_{d/2}(\kappa_s)}
     {I_{d/2-1}(\kappa_s)}$
where $I_\nu(\cdot)$ is the modified Bessel function of the first kind of order $\nu$~\cite{mardia2000directional}.
Mixture weights (speaker priors) $\pi_s$ are updated using type-II maximum likelihood~\cite{bishop2006pattern} as
\vspace{-1mm}
\begin{equation}
\label{eq:speaker_prior_update}
\textstyle
\pi_s \propto \sum_{t=1}^{T} \gamma_{ts}.
\end{equation}
\vspace{-1mm}
Inference alternates between updates
\eqref{eq:speaker_models_update}–\eqref{eq:speaker_prior_update}
until convergence.

\vspace{-0.5mm}
\section{Integration to EEND-VC}
\label{sec:Integrate_EEND-VC}

EEND-VC combines (i) a local EEND model that estimates speaker activity within short, overlapping windows with (ii) a global clustering stage that links these local speaker instances across the full recording. 
For each window, we extract one speaker embedding per active local speaker and pass the set of embeddings to the clustering stage, whose output is a consistent \emph{global} speaker identity for every local speaker instance.
Because local speakers within the same window represent distinct speakers, the second stage should satisfy a within-window \emph{``cannot-link''} constraint: two local speakers from the same window cannot be assigned the same global identity.

\begin{figure}[t]
  \centering

  \begin{subfigure}{\columnwidth}
    \centering

    \includegraphics[width=\linewidth]{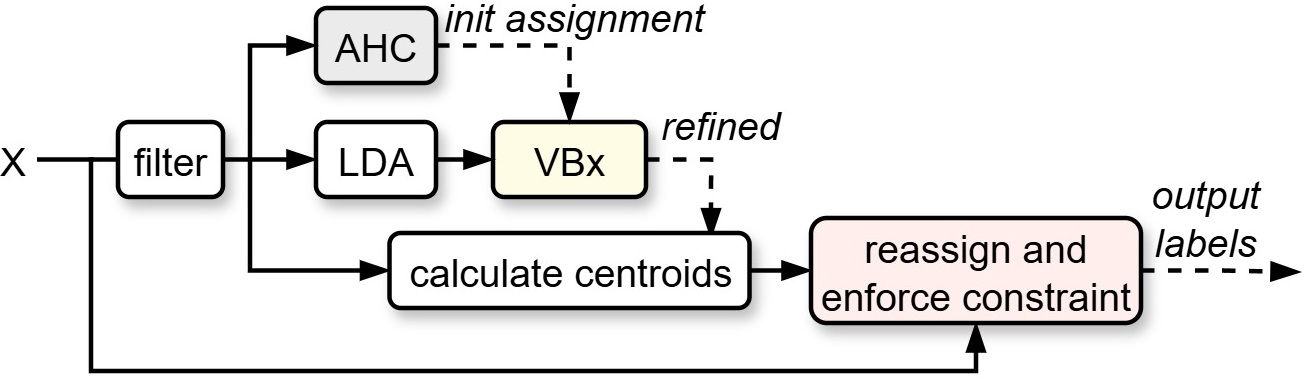}
    \caption[Baseline constrained PLDA VBx system.]%
    {Baseline VBx system~\cite{palka2025vbx} with heuristic short-embedding removal and post-hoc reassignment with enforcement of EEND-VC constraints.}
    \label{fig:pipeline:baseline}
  \end{subfigure}

   \vspace{0.3em} 

  \begin{subfigure}{\columnwidth}
    \centering
    \includegraphics[width=\linewidth]{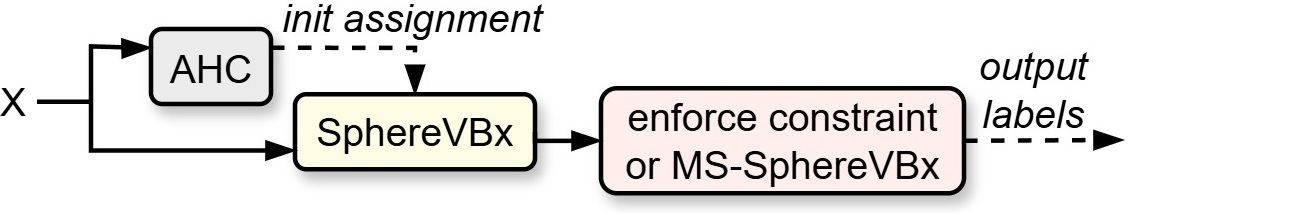}
    
        \caption{
    Simplified SphereVBx system that directly enforces constraint or uses Multi-Stream (MS) model extension. 
    }
    \label{fig:pipeline:proposed}
  \end{subfigure}
  \caption{Second stage of the EEND-VC pipeline. Dashed arrows represent labels. X denotes speaker embeddings.} 
  \label{fig:pipeline}
    \vspace{-4mm}
\end{figure}

Figure~\ref{fig:pipeline} summarizes two alternative realizations of this second stage; in both cases, an optional Agglomerative Hierarchical Clustering (AHC) can speed up inference by providing an initial clustering that is converted into hard responsibilities~\eqref{eq:speaker_responsibilities_update} for the first VB iteration. 
In the DiariZen baseline (Fig.~\ref{fig:pipeline:baseline})~\cite{palka2025vbx}, embeddings extracted from short segments ($<1.6\,\mathrm{s}$) are filtered out because they tend to be unreliable and can degrade clustering performance.
The remaining embeddings are processed by the standard VBx backend transform (global mean subtraction and length normalization, Linear Discriminant Analysis (LDA) projection for dimensionality reduction, centering, and length normalization) and clustered using PLDA-based VBx.
After VBx, global centroids are computed from the clustered embeddings and \emph{all} local embeddings are reassigned to these centroids using cosine similarity, following~\cite{palka2025vbx}. During this reassignment, the cannot-link constraint is enforced independently in each window through a one-to-one assignment between speakers within a local window and global centroids (illustration in~\cite{palka2025vbx}).
Overall, this baseline achieves strong performance, but relies on several decoupled heuristic processing blocks.

Our proposed integration (Fig.~\ref{fig:pipeline:proposed}) replaces VBx with SphereVBx, which models the original length-normalized embeddings using T-PSDA. Instead of discarding unreliable short embeddings, SphereVBx keeps \emph{all} embeddings in inference and reduces the impact of unreliable ones via duration-based weights as described in Section~\ref{subsec:weights}.
The cannot-link constraint can then be handled by the same per-window one-to-one assignment strategy as in the baseline system, using the inferred speaker responsibilities~\eqref{eq:speaker_responsibilities_update} instead of cosine similarity scores.
Alternatively, the constraint can be incorporated directly into the probabilistic model via the Multi-Stream (MS)-SphereVBx variant described in Section~\ref{subsec:MSVBX}.
As a result, SphereVBx yields a simpler and unified probabilistic second stage for EEND-VC.


\subsection{Reliability weights for observations}

\label{subsec:weights}

Instead of discarding unreliable short-segment embeddings as in the baseline, we keep all embeddings $\mathbf{x}_t$ and use reliability weights $w_t$ to reduce their influence on the clustering. This preserves direct assignment of every embedding to a global speaker, while avoiding the full-utterance centroid estimation and subsequent reassignment step used in the baseline (Fig.~\ref{fig:pipeline:baseline}). 
Concretely, we define weighted responsibilities as $\gamma'_{ts}=w_t\gamma_{ts}$ and use $\gamma'_{ts}$ in place of $\gamma_{ts}$ in the updates of the variational speaker posterior~\eqref{eq:speaker_models_update} and speaker prior~\eqref{eq:speaker_prior_update}, causing less reliable embeddings to contribute proportionally less to both.
In this work, the reliability weights are derived solely from segment duration. We use a simple binary weighting scheme with the same $1.6\,\mathrm{s}$ threshold that the DiariZen baseline uses for filtering short segments. More generally, the weights could be derived from other confidence cues or learned reliability. 

\subsection{Multi-Stream SphereVBx}
\label{subsec:MSVBX}
  Within a local EEND window, the detected speakers are distinct, so their embeddings must map to different global speakers, the cannot-link constraint.
  MS-VBx~\cite{delcroix2023multi} introduces this into VBx/SphereVBx by jointly assigning distinct global speakers to all local speakers in a window, rather than just scoring each embedding on its own.
    We refer the reader to~\cite{delcroix2023multi} for details of the model and focus here on an efficient implementation. 
    
    Consider a single local EEND window with $L$ extracted speaker embeddings. SphereVBx provides per-embedding, per-global-speaker emission log-scores $e_{\ell s}=\log \overline{p}(\mathbf{x}_\ell \mid s)$ defined in Eq.~\eqref{eq:expected_llh}, where $\ell$ and $s$ index embeddings within the window and global speakers, respectively.
    We form an $L$-way tensor $\boldsymbol{\Gamma}$ representing log-scores for all possible joint assignments of global speakers to the $L$ local speakers in the window. Each entry $\Gamma(s_1,\dots,s_L)$ corresponds to assigning global speaker $s_\ell$ to local speaker $\ell$ for all $\ell=1,\dots,L$ and equals $\sum_{\ell=1}^{L} e_{\ell s_\ell}$, i.e., the total score of the assignment $(s_1,\dots,s_L)$. The tensor is efficiently computed as an \emph{outer sum} of the row-wise scores $e_{\ell s}$.
    Assignments containing repeated global speaker indices are then excluded, which directly enforces the cannot-link constraint. After normalization, $\boldsymbol{\Gamma}$ yields posterior probabilities over the remaining valid assignments. The prior weights over valid permutation states are updated analogously to Eq.~\eqref{eq:speaker_prior_update}, now accumulating these permutation-state responsibilities in place of the per-speaker ones.
    

As in MS-VBx~\cite{delcroix2023multi}, the speaker models in Eq.~\eqref{eq:speaker_models_update} are estimated per global speaker rather than per permutation state, so they require speaker-level responsibilities. We recover these by marginalizing the normalized posterior $\boldsymbol{\Gamma}$: for each local embedding $\ell$ and global speaker $s$, we sum the probabilities of all valid states assigning $s$ to $\ell$. The resulting responsibilities 
are then plugged into Eq.~\eqref{eq:speaker_models_update}. This completes the MS inference.

\newcommand{\ccolspace}{0.75cm} 
\begin{table}[b]
\centering
\setlength{\tabcolsep}{3.5pt}
\caption{DER (\%) comparison across datasets for the cascade VAD+VBx+OSD diarization pipeline 
with * denoting values recomputed without a collar to match our evaluation.}
\vspace{-0.6em}

\resizebox{\linewidth}{!}{%
\label{tab:der_cascade_comparison}
\begin{tabular}{l | C{\ccolspace} C{\ccolspace} C{\ccolspace} C{\ccolspace} C{\ccolspace} C{\ccolspace} C{\ccolspace} | c }
\hline
Clustering & \textbf{AIS} & \textbf{AliM} & \textbf{AMI} & \textbf{DH3} & \textbf{MSD} & 
\makebox[0pt][c]{\textbf{RAMC}} & \makebox[0pt][c]{\textbf{VoxC}} &
Avg. \\
\hline
VBx \cite{landini2024diaper}
 & 15.8 & 28.8 & 34.6 & 20.3 & 29.5* & 18.2 & 11.1* & 22.6 \\
\hline
VBx (ours)
 & 16.7 & 29.0 & 35.5 & 20.7 & 27.7 & 18.4 & 10.7 & 22.7 \\
SphereVBx
 & 16.1 & 28.7 & 33.5 & 19.8 & 27.4 & 18.6 & 10.3 & 22.1\\
SphereVBx-PF
 & 16.0 & 28.8 & 33.9 & 19.9 & 27.9 & 18.5 & 10.4 & 22.2 \\ 
\hline
\end{tabular}
}
\vspace{-0.3cm}
\end{table}

\newcommand{\dshead}[4]{ 
  \makecell{\textbf{#1}\\ \small\color{darkgray!75}{ #3 / #4}} 
} 

\begin{table*}[!t]
\centering
\caption{Column headers indicate test sets statistics with average recording length in \textcolor{darkgray!75}{minutes} and speaker range \textcolor{darkgray!75}{min–max}. The table reports DER (\%) and, for baseline and proposed methods, mean speaker count error (MSCE) in parentheses. The last row reports the best results (SOTA) from previously 
published peer-reviewed works as of February 2026.}
\label{tab:clustering_sota}
\vspace{-0.3cm}
\resizebox{0.99\textwidth}{!}{
\begin{tabular}{l | 
c c c c c c c c
| c }
\toprule
System & 
\dshead{AIS}{20}{38.2}{5--7} &
\dshead{AliM}{20}{32.4}{2--4} &
\dshead{AMI}{16}{34.1}{3--4} &
\dshead{DH3}{259}{7.7}{1--9} &
\dshead{MSD}{490}{1.2}{2--4} &
\dshead{NSF}{160}{6.3}{3--7} &
\dshead{RAMC}{43}{28.8}{2--2} &
\dshead{VoxC}{232}{11.3}{1--21} &
Average \\

\midrule
Baseline & 9.9 & 10.8 & 13.9 & 14.5 & 15.7 & 16.7 & 11.0 & 8.8 & 12.65 (0.37)\\
\midrule
\midrule
SphereVBx        & 9.6 & 10.7 & 13.7 & 14.3 & 15.5 & 16.7 & 10.9 & 8.8 & 12.52 (0.34)\\ 
MS-SphereVBx     & 9.6 & 10.7 & 13.7 & 14.2 & 15.6 & 16.6 & 10.9 & 8.9 & 12.52 (0.32) \\
MS-SphereVBx-PF  & 9.7 & 10.6 & 13.7 & 14.1 & 15.8 & 16.5 & 10.6 & 8.9 & 12.48 (0.37) \\ 

\midrule
SOTA 02/2026  
& \makecell{ 9.8 \cite{han2025efficientgeneralizablespeakerdiarization}} 
& \makecell{10.8 \cite{han2025efficientgeneralizablespeakerdiarization}} 
& \makecell{13.9 \cite{han2025efficientgeneralizablespeakerdiarization}} 
& \makecell{14.5 \cite{han2025efficientgeneralizablespeakerdiarization}} 
& \makecell{15.6 \cite{han2025efficientgeneralizablespeakerdiarization}} 
& \makecell{16.7 \cite{han2025efficientgeneralizablespeakerdiarization}} 
& \makecell{10.3 \cite{broughton2025pushing}} 
& \makecell{ 8.6 \cite{liao2026dual}} 
& \makecell{-} \\ 

\bottomrule
\end{tabular}
}
\end{table*}

\vspace{-0.5mm}
\section{Experimental setup}
\label{sec:experimental_setup}

We evaluated on the following standard benchmark datasets: AMI~\cite{carletta2005ami}, \mbox{AISHELL-4} (AIS)~\cite{fu21b_interspeech}, \mbox{AliMeeting} (AliM)~\cite{yu2022m2met}, \mbox{NOTSOFAR-1 (NSF)}~\cite{vinnikov24_interspeech}, \mbox{MSDWild} (MSD)~\cite{liu22t_interspeech}, DIHARD3 full (DH3)~\cite{ryant21_interspeech}, RAMC~\cite{yang22h_interspeech}, and VoxConverse (VoxC)~\cite{chung20_interspeech}. For AMI and AliMeeting, we consistently use
the first channel from the far-field microphone array.
The cascade VBx setup (Sections~\ref{subsec:config_cascade}, \ref{subsec:comparision_cascade}) follows~\cite{landini2024diaper}, while the EEND-VC setup (Sections~\ref{subsec:config_EENDVC}, \ref{subsec:comparison_EENDVC}) uses the splits of~\cite{han2025efficientgeneralizablespeakerdiarization}. 
System performance is evaluated using diarization error rate (DER) without a forgiving collar. We also report macro-averaged DER across datasets and the mean speaker count error (MSCE), as defined in \cite{palka2025vbx}.

 For VBx, embeddings are projected by LDA from 256 to 128 dimensions, while SphereVBx uses loading matrix $\mathbf{K}$~\eqref{eq:TPSDA_P_W} to project embeddings to the 128 D subspace.

 \vspace{-2mm}



\subsection{Cascaded VAD + VBx + OSD pipeline configuration}
\label{subsec:config_cascade}

We first evaluate SphereVBx within a traditional cascaded diarization pipeline, where speaker embeddings extracted from short speech segments are clustered to produce a speaker segmentation of the recording. We use the same experimental setup as~\cite{landini2024diaper}, with their reported VBx results serving as the baseline. All components are kept fixed across experiments, including voice activity detection (VAD), overlapped speech detection (OSD), overlap handling, the ResNet101 embedding extractor, and AHC initialization. The SphereVBx variant differs from the baseline only in replacing PLDA-based VBx inference with SphereVBx, with $F_A$ and $F_B$ re-tuned accordingly.




\subsection{EEND-VC pipeline configuration}
\label{subsec:config_EENDVC}

We adopt the publicly available DiariZen \cite{han2025efficientgeneralizablespeakerdiarization} model\footnote{https://huggingface.co/BUT-FIT/diarizen-wavlm-large-s80-md-v2} as our baseline. 
The EEND component is based on a WavLM Large \cite{9814838} feature extractor followed by a Conformer encoder \cite{gulati20_interspeech}. 
The model was trained on $16$s chunks of the audio using the powerset loss \cite{plaquet23_interspeech}, which supports up to four speakers with arbitrary overlap. 
The pretrained model is kept fixed across all experiments and serves as the local EEND. 



For inference, we segment the audio into $16$s chunks with a shift of $1.6$s to perform local EEND as described in \cite{han2025leveraging} and extract speaker embeddings with the ResNet34-LM from the Wespeaker toolkit \cite{wang2023wespeaker}, trained on the VoxCeleb2 dataset \cite{chung18b_interspeech}, on which the PLDA and T-PSDA\footnote{https://github.com/bsxfan/Toroidal-PSDA} models were also trained.
Following \cite{palka2025vbx}, we merge speaker segments separated by silence gaps shorter than $\Delta = 0.5\,\mathrm{s}$ when evaluating on VoxConverse to match its annotation convention.


\section{Results}
\subsection{Comparison of VBx vs SphereVBx}
\label{subsec:comparision_cascade}

Table~\ref{tab:der_cascade_comparison} summarizes the performance of the cascade diarization pipeline (VAD + VBx + OSD)  under different speaker models for VBx and evaluation protocols. The first row reports the VBx baseline from \cite{landini2024diaper}. In that work, the hyperparameters $F_A$ and $F_B$ were tuned per dataset on its development partition.

In the remainder of the table, we evaluate all systems under a unified protocol used throughout this paper: we tune a single shared VBx hyperparameter configuration jointly on the full dev set and apply it across all test sets. Because the original PLDA training data processing and augmentation pipeline of \cite{landini2024diaper} is not available, we retrain the speaker models using our training setup to ensure a fair comparison under identical conditions.

The second row, VBx (ours), provides the reference system evaluated with VBx hyperparameters shared across all datasets. 
We show that a single configuration can serve as a practical and competitive operating point. Replacing VBx here with SphereVBx yields systematic gains, improving the average DER and reducing DER on most datasets. Finally, SphereVBx-PF achieves performance close to SphereVBx while eliminating the need for a pretrained model. 

 \subsection{Evaluation of SphereVBx in the EEND-VC pipeline}
\label{subsec:comparison_EENDVC}

Table~\ref{tab:clustering_sota} reports diarization performance of different SphereVBx variants integrated into the EEND-VC pipeline; the first row shows the VBx DiariZen baseline~\cite{han2025efficientgeneralizablespeakerdiarization}. 
The SphereVBx system enforces the cannot-link constraint using the responsibility-based per-window one-to-one assignment described in Section~\ref{sec:Integrate_EEND-VC}. Despite the substantially simplified clustering pipeline, DER is already improved compared to the strong baseline.

The number of global speakers inferred by clustering can still become smaller than the number of local speakers predicted by EEND within a window. Both the baseline and SphereVBx handle this by an ad-hoc mechanism: an auxiliary speaker label is retained to absorb embeddings that do not match any of the inferred global speakers. Removing this mechanism leads to an increase of about $1\%$ DER on the MSDWild dataset.

The MS-SphereVBx variant avoids this issue by enforcing the cannot-link constraint directly within probabilistic inference via the multi-stream formulation. As a result, it naturally maintains enough global speakers to explain the local EEND predictions. Its DER is comparable to constrained SphereVBx.

The MS-SphereVBx-PF variant achieves competitive performance without a pretrained T-PSDA, making it easier to deploy with different embedding extractors. To compensate for the absence of a trained T-PSDA backend, we set $F_A > 10$ and $F_B < 1$, unlike typical settings ($F_A < 1$, $F_A < F_B$).


Finally, the MSCE values, shown in parentheses in the last column, for all SphereVBx variants remain within the same range as the baseline VBx system. This indicates that the proposed clustering improves DER while maintaining accurate speaker-count estimation.

\vspace{-2mm}
\section{Conclusion}
\label{sec:conclusion}
We introduced SphereVBx, a general clustering method for normalized embedding vectors on the unit hypersphere using a T-PSDA speaker model within the VBx inference framework. SphereVBx and its parameter-free version allow a more principled and greatly simplified clustering pipeline by eliminating several heuristic steps required when applying standard VBx to EEND-VC. 
This simplified pipeline opens new possibilities such as E2E optimization of EEND-VC with SphereVBx~\cite{klement2024discriminative}, which we plan to explore next.  Future work will also apply SphereVBx beyond speaker diarization, for example, to face embeddings~\cite{deng2019arcface} or spatial features for source separation~\cite{Johannes_tsalp14}.
\section{Acknowledgments}
This work was supported by the project ``On our own: Opportunities and Risks in the Individualization of Society (PRINS) CZ.02.01.01/00/23\_025/0008710'', which is co-financed by the European Union. Computing on the IT4I supercomputer was supported by the Czech Ministry of Education, Youth and Sports through the e-INFRA CZ (ID:90254).

{\color{black}
\section{Generative AI Use Disclosure}
Generative AI tools (e.g., ChatGPT) were used solely for language editing and paraphrasing to improve the clarity and readability of the manuscript. The scientific content, methodology, experiments, and conclusions were developed by the authors. All authors take full responsibility for the accuracy, integrity, and originality of the work and consent to its submission.
}
\bibliographystyle{IEEEtran}
\bibliography{mybib}

@ARTICLE{Johannes_tsalp14,
  author={Traa, Johannes and Smaragdis, Paris},
  journal={IEEE/ACM Transactions on Audio, Speech, and Language Processing}, 
  title={Multichannel Source Separation and Tracking With RANSAC and Directional Statistics}, 
  year={2014},
  volume={22},
  number={12},
  pages={2233-2243},
  }

@article{han2025efficientgeneralizablespeakerdiarization,
    title={Efficient and Generalizable Speaker Diarization via Structured Pruning of Self-Supervised Models}, 
    author={Jiangyu Han and Petr Pálka and Marc Delcroix and Federico Landini and Johan Rohdin and Jan Cernocký and Lukáš Burget},
    journal={arXiv preprint arXiv:2506.18623, accepted by TASLP},
    year={2026},
}

@inproceedings{carletta2005ami,
  title={{The AMI meeting corpus: A pre-announcement}},
  author={Carletta, Jean and Ashby, Simone and Bourban, Sebastien and Flynn, Mike and Guillemot, Mael and Hain, Thomas and Kadlec, Jaroslav and Karaiskos, Vasilis and Kraaij, Wessel and Kronenthal, Melissa and others},
  booktitle={International workshop on machine learning for multimodal interaction},
  pages={28--39},
  year={2006},
  organization={Springer}
}

@InProceedings{deng2019arcface,
    author = {Deng, Jiankang and Guo, Jia and Xue, Niannan and Zafeiriou, Stefanos},
    title = {{ArcFace: Additive Angular Margin Loss for Deep Face Recognition}},
    booktitle = {{Proceedings of the IEEE/CVF Conference on Computer Vision and Pattern Recognition (CVPR)}},
    month = {June},
    year = {2019}
}

@inproceedings{klement2024discriminative,
  title={{Discriminative Training of VBx Diarization}},
  author={Klement, Dominik and Diez, Mireia and Landini, Federico and Burget, Luk{\'a}{\v{s}} and Silnova, Anna and Delcroix, Marc and Tawara, Naohiro},
  booktitle={Proc. ICASSP},
  pages={11871--11875},
  year={2024},
  organization={IEEE}
}

@article{landini2022bayesian,
  title={{Bayesian HMM Clustering of x-vector Sequences (VBx) in Speaker Diarization: Theory, Implementation and Analysis on Standard Tasks}},
  author={Landini, Federico and Profant, J{\'a}n and Diez, Mireia and Burget, Luk{\'a}{\v{s}}},
  journal={Computer Speech \& Language},
  volume={71},
  year={2022},
  publisher={Elsevier}
}

@article{landini2024diaper,
  title={{DiaPer}: End-to-End Neural Diarization with Perceiver-Based Attractors},
  author={Landini, Federico and Diez, Mireia and Stafylakis, Themos and Burget, Luk{\'a}{\v{s}}},
  journal={IEEE/ACM Transactions on Audio, Speech, and Language Processing},
  year={2024},
  publisher={IEEE}
}

@inproceedings{wang2023wespeaker,
  title={{Wespeaker: A research and production oriented speaker embedding learning toolkit}},
  author={Wang, Hongji and Liang, Chengdong and Wang, Shuai and Chen, Zhengyang and Zhang, Binbin and Xiang, Xu and Deng, Yanlei and Qian, Yanmin},
  booktitle={Proc. ICASSP},
  pages={1--5},
  year={2023},
  organization={IEEE}
}

@inproceedings{xiang2019margin,
  title={Margin matters: Towards more discriminative deep neural network embeddings for speaker recognition},
  author={Xiang, Xu and Wang, Shuai and Huang, Houjun and Qian, Yanmin and Yu, Kai},
  booktitle={2019 Asia-Pacific Signal and Information Processing Association Annual Summit and Conference (APSIPA ASC)},
  pages={1652--1656},
  year={2019},
  organization={IEEE}
}

@inproceedings{yu2022m2met,
  title={{M2MeT: The ICASSP 2022 multi-channel multi-party meeting transcription challenge}},
  author={Yu, Fan and Zhang, Shiliang and Fu, Yihui and Xie, Lei and Zheng, Siqi and Du, Zhihao and Huang, Weilong and Guo, Pengcheng and Yan, Zhijie and Ma, Bin and others},
  booktitle={Proc. ICASSP},
  pages={6167--6171},
  year={2022},
  organization={IEEE}
}

@book{bishop2006pattern,
  author    = {Christopher M. Bishop},
  title     = {Pattern Recognition and Machine Learning},
  series    = {Information Science and Statistics},
  publisher = {Springer},
  address   = {New York, NY},
  year      = {2006},
  edition   = {1},
  isbn      = {978-0-387-31073-2},
  doi       = {10.1007/978-0-387-45528-0}
}

@inproceedings{twO_covar_villalba11_interspeech,
  title     = {Towards fully Bayesian speaker recognition: integrating out the between-speaker covariance},
  author    = {Jesús Villalba and Niko Brümmer},
  year      = {2011},
  booktitle = {Proc. Interspeech},
  pages     = {505--508},
}

@inproceedings{diez19_interspeech,
  title     = {Bayesian HMM Based x-Vector Clustering for Speaker Diarization},
  author    = {Mireia Diez and Lukáš Burget and Shuai Wang and Johan Rohdin and Jan Černocký},
  year      = {2019},
  booktitle = {Interspeech 2019},
  pages     = {346--350},
  doi       = {10.21437/Interspeech.2019-2813},
  issn      = {2958-1796},
}

@inproceedings{brummer10_odyssey,
  title={The speaker partitioning problem.},
  author={Br{\"u}mmer, Niko and De Villiers, Edward},
  booktitle={Odyssey},
  pages={34},
  year={2010}
}

@inproceedings{chung18b_interspeech,
  title     = {{VoxCeleb2}: Deep Speaker Recognition},
  author    = {Joon Son Chung and Arsha Nagrani and Andrew Zisserman},
  year      = {2018},
  booktitle = {Proc. Interspeech},
  pages     = {1086--1090},
}

@ARTICLE{9814838,
  author={Chen, Sanyuan and Wang, Chengyi and Chen, Zhengyang and Wu, Yu and Liu, Shujie and Chen, Zhuo and Li, Jinyu and Kanda, Naoyuki and Yoshioka, Takuya and Xiao, Xiong and Wu, Jian and Zhou, Long and Ren, Shuo and Qian, Yanmin and Qian, Yao and Wu, Jian and Zeng, Michael and Yu, Xiangzhan and Wei, Furu},
  journal={IEEE Journal of Selected Topics in Signal Processing}, 
  title={{WavLM}: Large-Scale Self-Supervised Pre-Training for Full Stack Speech Processing}, 
  year={2022},
  volume={16},
  number={6},
  pages={1505-1518},
  doi={10.1109/JSTSP.2022.3188113}}

@inproceedings{gulati20_interspeech,
  title     = {Conformer: Convolution-augmented Transformer for Speech Recognition},
  author    = {Anmol Gulati and James Qin and Chung-Cheng Chiu and Niki Parmar and Yu Zhang and Jiahui Yu and Wei Han and Shibo Wang and Zhengdong Zhang and Yonghui Wu and Ruoming Pang},
  year      = {2020},
  booktitle = {Proc. Interspeech},
  pages     = {5036--5040},
  doi       = {10.21437/Interspeech.2020-3015},
  issn      = {2958-1796},
}

@inproceedings{bredin23_interspeech,
  title     = {pyannote.audio 2.1 speaker diarization pipeline: principle, benchmark, and recipe},
  author    = {Hervé Bredin},
  year      = {2023},
  booktitle = {Proc. Interspeech},
  pages     = {1983--1987},
}

@inproceedings{plaquet23_interspeech,
  title     = {Powerset multi-class cross entropy loss for neural speaker diarization},
  author    = {Alexis Plaquet and Hervé Bredin},
  year      = {2023},
  booktitle = {Proc. Interspeech},
  pages     = {3222--3226},
}

@inproceedings{advances_kino,
  author= {Keisuke Kinoshita and Marc Delcroix and Naohiro Tawara\ },
  title     = {Advances in Integration of End-to-End Neural and Clustering-Based Diarization for Real Conversational Speech},
  year      = {2021},
  booktitle = {Interspeech 2021},
  pages     = {3565--3569},
  doi       = {10.21437/Interspeech.2021-1004},
  issn      = {2958-1796},
}

@inproceedings{kinoshita2021integrating,
  author={Kinoshita, Keisuke and Delcroix, Marc and Tawara, Naohiro},
  title={Integrating end-to-end neural and clustering-based diarization: Getting the best of both worlds},
  booktitle={Proc. ICASSP},
  pages={7198--7202},
  year={2021},
  organization={IEEE}
}

@inproceedings{fu21b_interspeech, 
  title     = {{AISHELL-4}: An Open Source Dataset for Speech Enhancement, Separation, Recognition and Speaker Diarization in Conference Scenario},
  author    = {Yihui Fu and Luyao Cheng and Shubo Lv and Yukai Jv and Yuxiang Kong and Zhuo Chen and Yanxin Hu and Lei Xie and Jian Wu and Hui Bu and Xin Xu and Jun Du and Jingdong Chen},
  year      = {2021},
  booktitle = {Proc. Interspeech},
  pages     = {3665--3669},
}

@inproceedings{ryant21_interspeech,
  title     = {The Third {DIHARD} Diarization Challenge},
  author    = {Neville Ryant and Prachi Singh and Venkat Krishnamohan and Rajat Varma and Kenneth Church and Christopher Cieri and Jun Du and Sriram Ganapathy and Mark Liberman},
  year      = {2021},
  booktitle = {Proc. Interspeech},
  pages     = {3570--3574},
}

@inproceedings{vinnikov24_interspeech,
  title     = {{NOTSOFAR-1} Challenge: New Datasets, Baseline, and Tasks for Distant Meeting Transcription},
  author    = {Vinnikov, Alon and Ivry, Amir and Hurvitz, Aviv and Abramovski, Igor and Koubi, Sharon and Gurvich, Ilya and Peer, Shai and Xiao, Xiong and Elizalde, Benjamin Martinez and Kanda, Naoyuki and others},
  year      = {2024},
  booktitle = {Proc. Interspeech},
  pages     = {5003--5007},
}

@inproceedings{liu22t_interspeech, 
  title     = {{MSDWild}: Multi-modal Speaker Diarization Dataset in the Wild},
  author    = {Tao Liu and Shuai Fan and Xu Xiang and Hongbo Song and Shaoxiong Lin and Jiaqi Sun and Tianyuan Han and Siyuan Chen and Binwei Yao and Sen Liu and Yifei Wu and Yanmin Qian and Kai Yu},
  year      = {2022},
  booktitle = {Proc. Interspeech},
  pages     = {1476--1480},
}

@inproceedings{yang22h_interspeech,
  title     = {Open Source {MagicData-RAMC}: A Rich Annotated Mandarin Conversational ({RAMC}) Speech Dataset},
  author    = {Zehui Yang and Yifan Chen and Lei Luo and Runyan Yang and Lingxuan Ye and Gaofeng Cheng and Ji Xu and Yaohui Jin and Qingqing Zhang and Pengyuan Zhang and Lei Xie and Yonghong Yan},
  year      = {2022},
  booktitle = {Proc. Interspeech},
  pages     = {1736--1740},
}

@inproceedings{chung20_interspeech,
  title     = {Spot the Conversation: Speaker Diarisation in the Wild},
  author    = {Joon Son Chung and Jaesung Huh and Arsha Nagrani and Triantafyllos Afouras and Andrew Zisserman},
  year      = {2020},
  booktitle = {Proc. Interspeech},
  pages     = {299--303},
}

@inproceedings{han2025leveraging,
  title={Leveraging self-supervised learning for speaker diarization},
  author={Han, Jiangyu and Landini, Federico and Rohdin, Johan and Silnova, Anna and Diez, Mireia and Burget, Luk{\'a}{\v{s}}},
  booktitle={Proc. ICASSP},
  pages={1--5},
  year={2025},
  organization={IEEE}
}

@inproceedings{delcroix2023multi,
  title={{Multi-stream extension of variational bayesian HMM clustering (MS-VBx) for combined end-to-end and vector clustering-based diarization}},
  author={Delcroix, M. and Tawara, N. and Diez, M. and others},
  booktitle={Proc. Interspeech},
  pages={3477--3481},
  year={2023}
}

@book{mardia2000directional,
  title={Directional Statistics},
  author={Mardia, K.V. and Jupp, P.E.},
  isbn={9780471953333},
  lccn={99033679},
  series={Wiley Series in Probability and Statistics},
  url={https://books.google.cz/books?id=zjPvAAAAMAAJ},
  year={2000},
  publisher={Wiley}
}

@techreport{diez2021vbx,
  author       = {Diez, Mireia and Burget, Luk{\'a}{\v{s}} and Federico, Landini},
  title        = {{CSL VBx Derivations}},
  institution  = {Brno University of Technology},
  year         = {2021},
  url          = {https://www.fit.vut.cz/person/mireia/public/CSL_VBHMM_tech_report.pdf}
}

@inproceedings{PSDA_brummer22_interspeech,
  title     = {{Probabilistic Spherical Discriminant Analysis: An Alternative to PLDA for length-normalized embeddings}},
  author    = {Niko Brummer and Albert Swart and Ladislav Mosner and Anna Silnova and Oldrich Plchot and Themos Stafylakis and Lukas Burget},
  year      = {2022},
  booktitle = {{Proc. Interspeech}},
  pages     = {1446--1450},
}

@INPROCEEDINGS{TPSDA_10095580,
  author={Silnova, Anna and Brümmer, Niko and Swart, Albert and Burget, Lukáš},
  booktitle={Proc. ICASSP}, 
  title={Toroidal Probabilistic Spherical Discriminant Analysis}, 
  year={2023},
  organization={IEEE}
}

@article{palka2025vbx,
  title={{VBx} for End-to-End Neural and Clustering-based Diarization},
  author={P{\'a}lka, Petr and Han, Jiangyu and Delcroix, Marc and Tawara, Naohiro and Burget, Luk{\'a}{\v{s}}},
  journal={arXiv preprint arXiv:2510.19572},
  year={2025}
}

@inproceedings{broughton2025pushing,
  title={Pushing the Limits of End-to-End Diarization},
  author={Broughton, S. J. and Samarakoon, L.},
  booktitle={Proc. Interspeech},
  pages={5218--5222},
  year={2025}
}

@article{liao2026dual,
  title={Dual-Strategy-Enhanced {ConBiMamba} for Neural Speaker Diarization},
  author={Liao, Zhen and Dai, Gaole and Chen, Mengqiao and Cheng, Wenqing and Xu, Wei},
  journal={arXiv preprint arXiv:2601.19472, accepted by ICASSP},
  year={2026}
}

@article{banerjee2005clustering,
  title={Clustering on the Unit Hypersphere using von Mises-Fisher Distributions.},
  author={Banerjee, Arindam and Dhillon, Inderjit S and Ghosh, Joydeep and Sra, Suvrit and Ridgeway, Greg},
  journal={Journal of Machine Learning Research},
  volume={6},
  number={9},
  year={2005}
}

@inproceedings{gopal2014mises,
  title={Von mises-fisher clustering models},
  author={Gopal, Siddharth and Yang, Yiming},
  booktitle={International conference on machine learning},
  pages={154--162},
  year={2014},
  organization={PMLR}
}

@inproceedings{dubey2018robust,
  title={Robust speaker clustering using mixtures of von mises-fisher distributions for naturalistic audio streams},
  author={Dubey, Harishchandra and Sangwan, Abhijeet and Hansen, John HL},
  booktitle={Proc. Interspeech},
  pages={3603--3607},
  year={2018}
}

@inproceedings{sholokhov2023,
  author={Sholokhov, Alexey and Kuzmin, Nikita and Lee, Kong Aik and Chng, Eng Siong},
  booktitle={Proc. ICASSP},
  title={Probabilistic Back-ends for Online Speaker Recognition and Clustering}, 
  year={2023},
  volume={},
  number={},
  }

\end{document}